\documentclass[prb,twocolumn,amssymb,showpacs]{revtex4}
\usepackage{graphicx,amsmath,mathrsfs}
\begin{document}

\title{Quantum Hall Effect from the Topological Surface States of Strained Bulk HgTe}

\author{C.\ Br\"une$^{1}$, C.X.\ Liu$^1$, E.G.\ Novik$^1$, E.M. Hankiewicz$^{1}$, H.\ Buhmann$^1$, Y.L. Chen$^2$, X.L. Qi$^2$, Z.X. Shen$^2$, S.C. Zhang$^2$ and L.W.\ Molenkamp$^1$ }

\affiliation{ $^1$Faculty for Physics and Astronomy and R\"{o}ntgen Center for Complex Material Systems, Universit\"at W\"urzburg, Am Hubland, D-97074, W\"urzburg, Germany\\
 $^2$Department of Physics, McCullough Building,
Stanford University
, Stanford, CA 94305-4045 }

\date\today

\begin{abstract}
We report transport studies on a three dimensional, 70 nm thick HgTe layer, which is strained by epitaxial growth on a CdTe substrate. The strain induces a band gap in the otherwise semi-metallic HgTe, which thus becomes a
three dimensional topological insulator. Contributions from residual bulk carriers to
the transport properties of the gapped HgTe layer are negligible at mK temperatures. As a result, the sample exhibits
a quantized Hall effect that results from the 2D single cone Dirac-like topological surface states.
\end{abstract}


\maketitle

The discovery of two (2D) and three dimensional (3D) topological insulators
(TI)\cite{bernevig2006c,kane2005A,koenig2007,fu2007a,fu2007b,moore2007,hsieh2008,zhang2009,xia2009,chen2009}
has generated strong activity in the condensed matter physics
community\cite{hasan2010,moore2010}.
Current research on 3D TIs is mostly focused on Bi$_2$Te$_3$, Bi$_2$Se$_3$ and Sb$_2$Te$_3$
compounds\cite{zhang2009,xia2009,chen2009} due to their simple Dirac-like surface states,
which have been observed by angle resolved photoemission spectroscopy (ARPES) and scanning tunneling
microscopy\cite{hasan2010}. However, these compounds show strong defect doping and low carrier mobility,
and the observation of surface charge transport is obscured by the bulk conductivity.
Many of the predicted novel properties of a 3D TI, such as the quantized magneto-electric
effect\cite{qi2008B,essin2009} and the surface Majorana fermions\cite{fu2008}, can only be observed when
bulk carriers are negligible compared to the surface states. Experimentally reaching the intrinsic TI regime,
where bulk carriers are absent, is now the central focus of the field.

The two dimensional TI state was first predicted and observed in 2D HgTe quantum wells (QW)\cite{bernevig2006c,koenig2007}, and non-local transport measurements demonstrate edge state transport without any contributions from 2D bulk carriers.\cite{roth2009} 3D HgTe is a semi-metal which is charge-neutral when the Fermi energy is at the touching point between the light-hole and heavy-hole $\Gamma_8$ bands at the Brillouin zone center. A unique property of the band structure of HgTe is the energetic inversion of the $\Gamma_6$ and $\Gamma_8$ band ordering, which is the origin of the quantum spin Hall effect in 2D HgTe/CdTe QWs\cite{koenig2007}. Due to the band inversion, 3D HgTe is also expected to have Dirac-like surface states\cite{Chang85,Pank90}, but since the material is semi-metallic, this state is always coupled to metallic bulk states. With applied strain, a gap opens up between the light-hole and heavy-hole bands, so that strained 3D HgTe is expected to be a 3D TI\cite{dai2008,fu2007a}. In this paper we demonstrate experimentally that a gap opens up in in-plane strained 3D HgTe bulk layers grown by  molecular beam epitaxy (MBE), and we reach the much sought after intrinsic TI regime in a material with negligible bulk carriers. In this regime, the Hall effect of the 3D HgTe bulk layer is quantized, due to the contributions from the surface states only. Theoretical considerations are in agreement with the experimental results and confirm the transport through 2D surface states with Dirac type dispersion.\\


HgTe bulk samples have been grown by MBE on CdTe subtrates, which have a lattice constant that is 0.3 \% larger than that of bulk HgTe (0.646 nm). At this mismatch, the critical thickness for lattice relaxation is around 200 nm, implying that for thinner HgTe the epilayer adopts the lateral lattice constant of the substrate, while in thicker layers the strain is relaxed by the formation of dislocations.

To provide evidence for the occurrence of the topological surface state, we first show an ARPES measurement on a 1~$\mu$m thick HgTe layer in  [Fig.~1 (a)]. In this layer, the lattice strain is fully relaxed, and the surface has the lattice constant of unstrained bulk material. The figure clearly shows the presence of the predominantly linearly dispersing surface state band (SSB), coexisting with bulk bands (BBs) (more data on these assignments can be found in the supplementary material). According to the theoretical analysis in Ref. (19),
the surface states originate from the inversion
between the $\Gamma_8$ light-hole and $\Gamma_6$ electron bands, while the bulk bands, which appear nearly at the same energy range with the surface states, correspond to the $\Gamma_8$ heavy-hole band.

Since the relaxed sample is a semi-metal and thus not a TI in the strict sense, we have grown a thinner sample, which is only 70 nm thick, thin enough for the epitaxial strain due to the lattice mismatch with the substrate to coherently strain the sample, thus opening up a bulk insulating gap. Fig.~1 (b) shows a high resolution X-ray diffraction map of the [115] reflex of this sample in units of reciprocal lattice space vectors.
The bright spot in the center of this graph is the reflex from the substrate, while the thin vertical
line stems from the HgTe epilayer. The absence of any relaxation of the reciprocal lattice vector $\bf{ Q_x}$ for the epilayer is direct evidence that this sample is fully strained.\cite{bauer1996} The deformation potentials
of (Hg,Cd)Te have been reported in the literature\cite{Adachi09}; using these values and the 0.3 \% lattice mismatch we calculate an energy gap of the order of $\sim 22$ meV for fully strained HgTe on
CdTe, using an eight band \textbf{k}$\cdot$\textbf{p} model \cite{novik2005}.

For transport experiments, we have subsequently patterned parts of both wafers into Hall-geometry devices with a mesa of  200 $\mu$m width and 600 $\mu$m length, using argon ion etching. The magneto transport of the samples has been investigated at a base temperature of 50 mK, in magnetic fields up to 16 T. In the 1~$\mu$m thick sample, the Hall data indicates that bulk conductance dominates the transport. Fig. 1 (c) shows that for this sample we observe a non-monotonic dependence of the Hall voltage, which is characteristic for the multi-carrier transport expected from a semi-metal. Much more interesting behavior is observed for the 70 nm-thick sample, of which the longitudinal and Hall resistance are shown in the inset of Fig.~2. From the low field data, the electron mobility can be extracted, and estimated as 34000 cm$^2$/(V$\cdot$s), which is significantly higher than that observed in Bi$_2$Se$_3$ and Bi$_2$Te$_3$\cite{analytis2010b,qu2010}. At high magnetic fields, the longitudinal and Hall resistance exhibit distinct features which are characteristic for a 2D electron system (2DES): the Hall resistance $R_{xy}$ shows plateaus at the same magnetic fields where the longitudinal resistance $R_{xx}$ develops a minimum (inset of Fig.~2). Additionally, the Hall resistance $R_{xy}$ shows the expected 2D quantized plateau values, which become clearer in a conductivity plot (Fig.~2).

Compared with the quantum Hall effect of an ordinary 2DES, two unusual observations should be emphasized, which indicate that the observed quantum Hall plateaus indeed result from the Dirac-type dispersion
of the topological surface states. First, at low magnetic fields, a sequence of Hall
plateaus develops with odd filling factors, $\nu$ = 9, 7,
and 5, before at higher field the sequence is continued with $\nu$ = 4, 3, and 2.
The occurrence of odd-number filling factors at low magnetic field indicates
the presence of a zero mode (a Landau level at zero energy) due to the
linear dispersion of Dirac fermions, as has also been found in graphene\cite{castro2009}
and in HgTe/CdTe QWs with a critical thickness of 6.3 nm\cite{buettner2010}.
A 70nm thick layer can safely be regarded as a 3D material; the confinement energies of bulk carriers
are sufficiently small that Hall quantization effects would be washed out by multi-subband averaging.
We thus assume that the Hall plateaus result from the topological surface states, and
use a model with two Dirac cones, one on the top (vacuum) and one on the bottom
(CdTe interface) to describe the system (see additional online material for details).
Because massless Dirac fermions in a magnetic field always exhibit a zero mode,
the Hall conductance of a single Dirac cone is given by $\sigma_{xy}=(n+\frac{1}{2}) e^2/h$,
where $n=n_t$ or $n=n_b$ (for the top and bottom surfaces, respectively)
is always an integer\cite{fu2007a,qi2008B}. The fractional factor of $\frac{1}{2}$
in the Hall conductance $\sigma_{xy}$ is a consequence of the quantized
bulk topological term in the electromagnetic action\cite{qi2008B},
and is independent of the microscopic details. When top and the bottom
surface have the same filling factor, {\it i.e.} $n=n_t=n_b$, the total
Hall conductance is given by $\sigma^T_{xy}=(2n+1) e^2/h$.
Therefore, within the two Dirac cone model, the odd filling factor
at low magnetic fields can be naturally explained assuming both
surfaces have the same density. The appearance of an even filling factor
at high magnetic field indicates that the degeneracy is removed.
In an ordinary 2D electron gas, such a lift of degeneracy may occur due to
Zeeman coupling. However, if a system with top and bottom surface states has inversion symmetry,
Zeeman coupling cannot split the energies of the two remote Dirac cones. As we explain in the supplementary material, this is because the inversion symmetry is preserved by the magnetic field. The
Landau levels from the top and bottom surface states will remain degenerate as long as
the hybridization between the two surface states is negligible. For a thickness of 70nm, hybridization between the
top and bottom surfaces can be neglected since the surface state width is around 2$\sim$3nm.
Thus we conclude that the inversion symmetry breaking is necessary for the explanation of the
Landau level splitting. In practice, the different electrostatic environments of both surfaces break the inversion symmetry and lead to different densities at both surfaces.
This then leads to different Landau filling factors for the top and bottom surfaces at high magnetic fields, and the
visibility of even filling factors at high fields results from the increased energy splitting between adjacent
Landau levels at lower filling factors in a Dirac system, as schematically indicated in the inset of Fig. 3.
A calculation of the density of states in a magnetic field from a two Dirac cones model with an inversion breaking term indeed agrees well with the minima of the Shubnikov-de Haas oscillation, as shown in Fig. 3 (see additional online material for details). From the fit, we obtain very reasonable carrier densities of  $3.7\times10^{11}$cm$^{-2}$ for the bottom (CdTe interface) and $4.8\times10^{11}$cm$^{-2}$ for the top surfaces, respectively. We note that the carrier densities found here also imply we can neglect any effects from bulk carriers - if the amount carriers found here would be evenly distributed over the 70 nm slab this would result in an overall 2D density of $\sim 10^{10}cm^{-2}$, yielding a very different quantum Hall behavior.

The second observation is that the minima in $R_{xx}$ do not approach zero even at the highest magnetic fields although the Hall resistance is quantized and the plateaus have the expected resistance value. This indicates that besides the chiral edge modes from the quantum Hall effect, there are other modes contributing to the longitudinal, but not the Hall transport. A plausible candidate are the metallic states at the side surfaces\cite{chu2010}. While the above two Dirac cone model only takes the top and bottom surfaces explicitly into account, the topological surface states also occur on the four side surfaces, which see a parallel, rather than perpendicular, magnetic field.
Consequently, at the side surfaces the Dirac points are not gapped, but only shifted by the applied magnetic field. Thus, the surface states at the side surfaces will remain metallic in magnetic field and coexist with the chiral edge states, as shown schematically in Fig 1(d). This provides a backscattering mechanism when the transport on the top and bottom surfaces is in the quantum Hall regime\cite{chu2010}. A more systematic study of the influence of the residual bulk conductivity and side surface states is required to fully understand the quantitative behavior of $R_{xx}$ and $R_{xy}$, which is beyond the scope of the present letter.

In summary, we have experimentally reached the much sought after regime of the intrinsic 3D topological insulator with negligible bulk carriers, in epitaxially strained 3D HgTe sample. Our observation of a quantized Hall conductance in a 3D sample conclusively demonstrates a key feature of 3D topological insulators. A simple model with two Dirac cones is proposed to understand the most salient features of the transport measurement qualitatively. The quality of our sample should be sufficient to observe the quantized topological magneto-electric effect\cite{qi2008B}, and directly determine the 3D topological invariant experimentally.

 We acknowledge useful discussions with G. Astakhov, J.E. Moore and A.H. MacDonald. This work was supported by
the German Research Foundation DFG (SPP 1285 Halbleiter Spintronik, DFG-JST joint research program,
and grants AS327/2-2 (E.G.N.) and HA5893/1-1(E.M.H.)), the Alexander von Humboldt Foundation (C.X.L. and S.C.Z.), the EU ERC-AG program (L.W.M.) and the US Department of Energy, Office of Basic Energy Sciences, Division of Materials Sciences and Engineering, under contract DE-AC02-76SF00515 (S.C.Z).

\begin{figure}
   \centering
   \includegraphics[width=0.4 \textwidth]{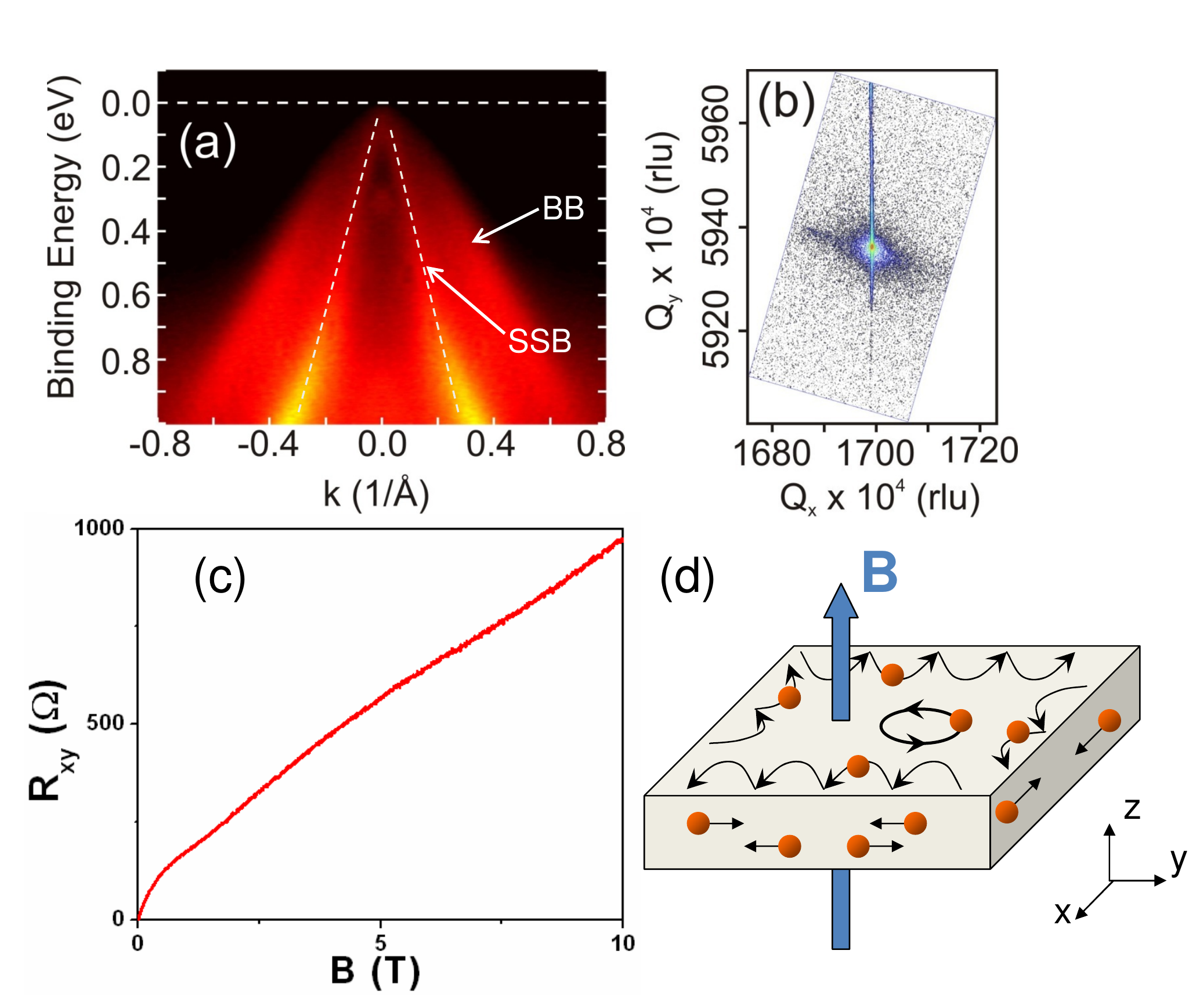}
    \caption{ {\bf Supporting characterization experiments}. (a) ARPES
measurements on a relaxed, 1~$\mu$m thick HgTe sample. The dispersion of the surface
state (SSB) and bulk bands (BB) are indicated by the arrows; (b) Reciprocal space
map in the region of the [115] reflex of the 70 nm thick HgTe sample demonstrating that the epilayer is coherently strained; (c) Hall resistance as a function of magnetic field of the semi-metallic 1~$\mu$m thick sample ; (d)
Schematic picture of the coexistence of the chiral edge states from the upper and lower surfaces and the non-chiral metallic modes at the side surfaces. The magnetic field is perpendicular to the upper and lower surfaces, but parallel to the four side surfaces. }
    \label{fig1}
\end{figure}

\begin{figure}
    \centering
    \includegraphics[width=0.4\textwidth]{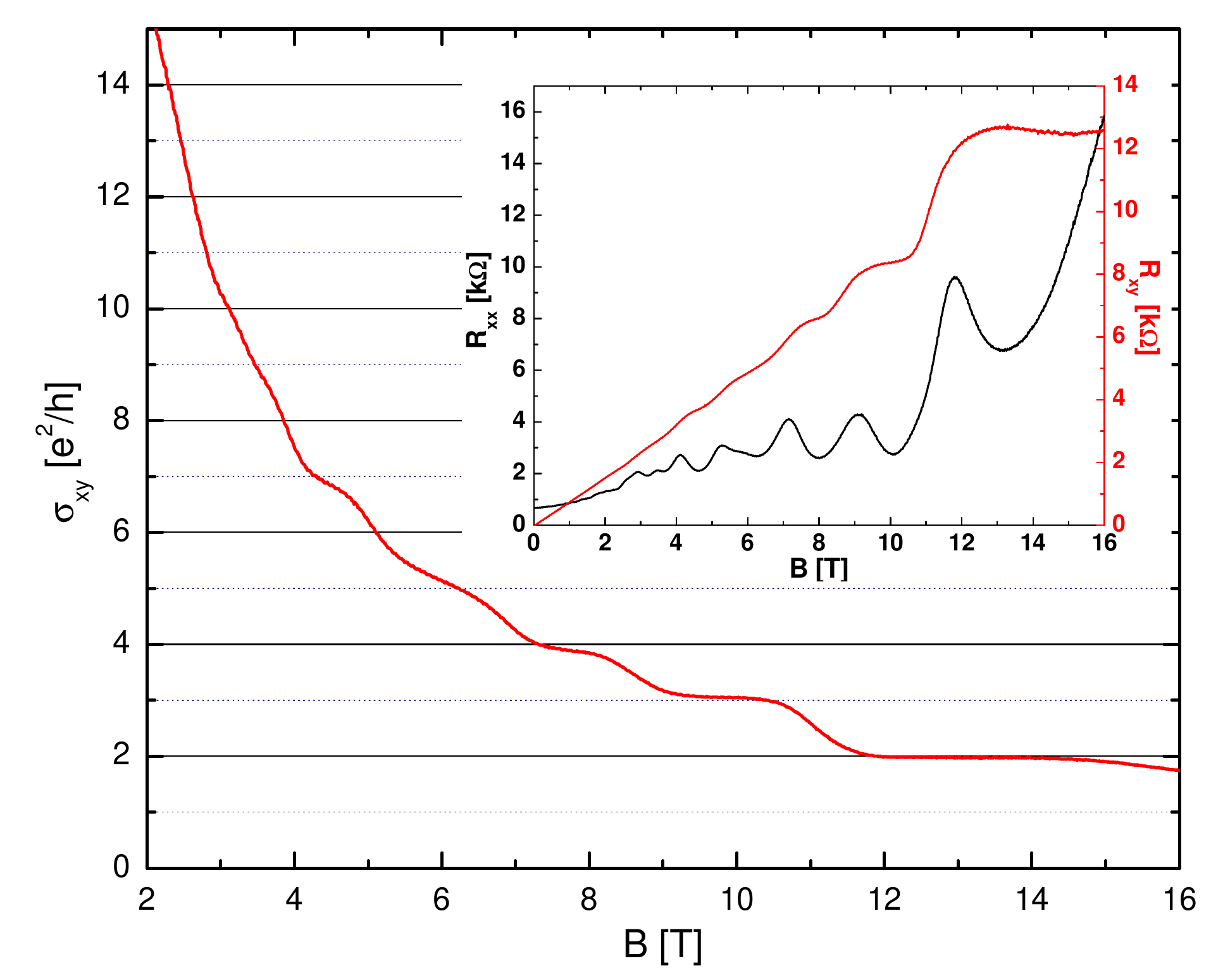}
    \caption{ {\bf Transport data on the strain-gapped 70 nm thick HgTe sample. } The Hall conductivity of the 70 nm thick HgTe sample measured at 50 mK shows plateaus at the quantized values. The inset shows the Hall resistance $R_{xy}$, together with the longitudinal resistance $R_{xx}$.}
    \label{fig2}
\end{figure}

\appendix

\section{Details of the ARPES Experiment}

Angle resolved photoemission spectroscopy (ARPES) experiments on
1$\mu m$ thick HgTe thin film (grown on CdTe substrate) were
performed at Beamline 10.0.1 of the Advanced Light Source (ALS).
During the experiments, the measurement chamber pressure was kept $<
4\times 10^{-11}$ Torr, and data were recorded by a Scienta R4000
analyzers at 15 K sample temperature. The photon energy used was 65
eV and the total convolved energy and angle resolutions were 20 meV
and 0.2 degree, respectively. The fresh surface for ARPES
measurement was obtained by gently sputtering the HgTe (001) surface
with Ar ion-beam, and no observable surface degradation was noticed
during typical experimental period (12 hrs). We also measured the
photon energy dependent (65-70eV) band dispersion of HgTe, as shown
in Fig \ref{ARPES}. For all measurements at different energies, the
surface state band can clearly be observed,
superimposed on a blurry background of bulk bands. The shape and dispersion of the
band do not vary with photon energy, while its intensity may vary. This which is direct evidence that
the feature does not have $k_z$ dispersion, as expected from a two dimensional (surface) state.
We note that at the vicinity of the Gamma point, the surface state band has gained finite curvature, similar to that observed\cite{xia2009,chen2010c} in Bi$_2$Se$_3$. The difference between the HgTe and the Bi2Se3 cases is that for the latter, both the upper and lower Dirac cone can be seen, while for the former case, the Dirac point resides close to the Fermi energy and thus only the lower Dirac cone can seen in the ARPES measurements of Fig. \ref{ARPES}.

\begin{figure}
    \centering
    \includegraphics[width=0.4\textwidth]{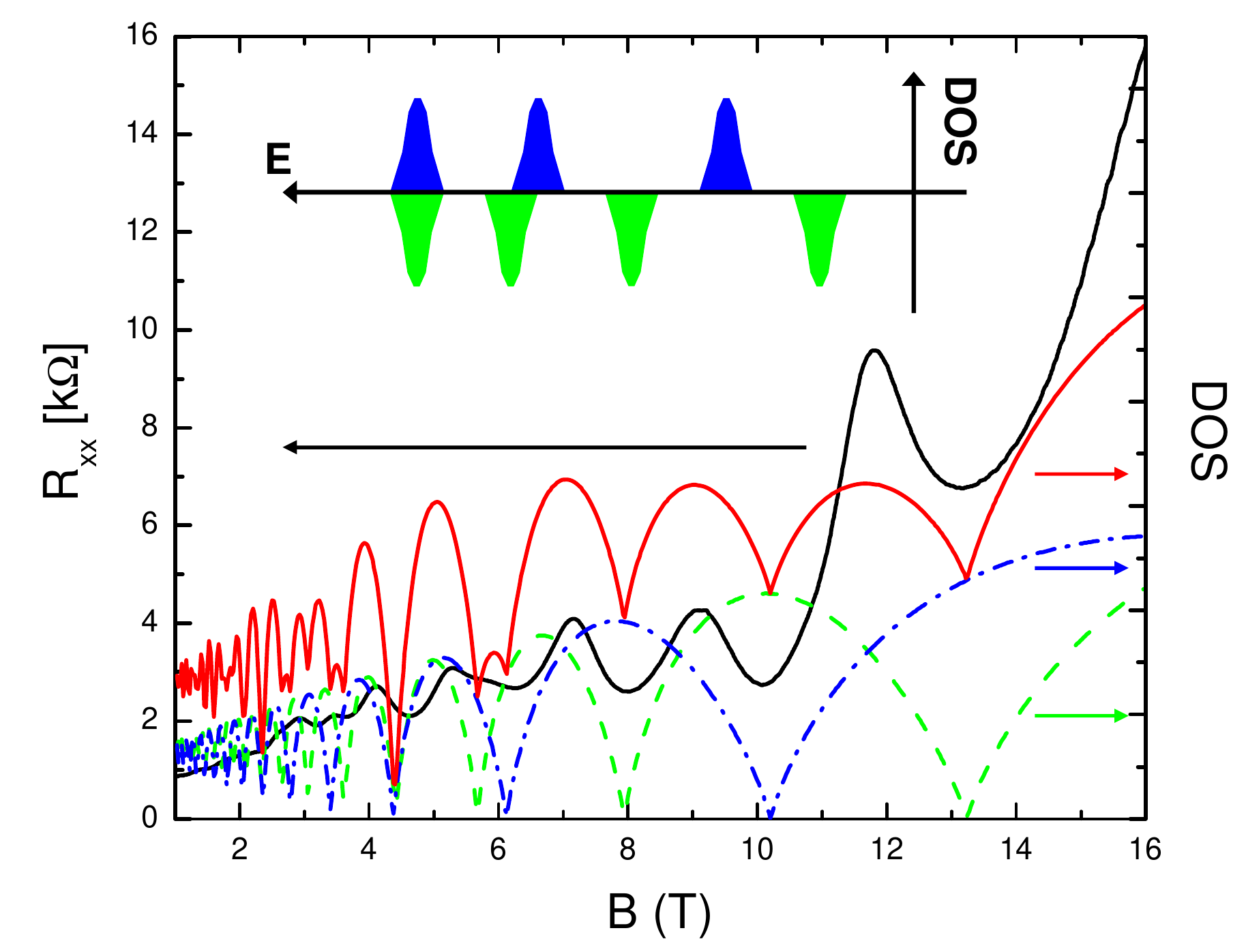}
    \caption{{\bf Comparison between the calculated density of states (DOS) and the measured Shubnikov- de Haas oscillations.} The DOS are calculated for a two Dirac cones model for the surface states with one Dirac cone (at the CdTe interface) having a carrier density $3.7\times10^{11}$cm$^{-2}$ (blue dashed-dotted line) and the other (the surface facing vacuum) $4.8\times10^{11}$cm$^{-2}$ (green dashed line). The sum of the DOS for the two Dirac cones (red line) compares well with the measured longitudinal resistance $R_{xx}$ (black line). The inset shows the Landau ladders of the Dirac fermions on both surfaces.}
    \label{fig3}
\end{figure}

\section{ k$ \bullet $p calculation of the band structure of strained ${\rm HgTe}$}

To understand the transport properties of the 70nm sample, we first
calculate the band dispersion of the structure using the ${\bf k} \bullet {\bf p}$ type approach.
HgTe can be well described by the eight-band Kane model, where the electron $\Gamma_6$ band and hole $\Gamma_7$, $\Gamma_8$ bands are taken into account explicitly. The detailed form of the Kane model, as well as the related band parameters, can be found in Ref \onlinecite{novik2005}. As is evident from the reciprocal lattice map of Fig. 1b in the main article, the 70 nm thick HgTe epilayer is fully strained due to the lattice mismatch between CdTe substrate and HgTe. Although this lattice mismatch is small
($\epsilon=(a_{CdTe}-a_{HgTe})/a_{HgTe}\approx0.003$ with $a_{CdTe}$ and $a_{HgTe}$ lattice constants
of CdTe and HgTe, respectively), the resulting strain is sufficient to open a bandgap between
the heavy- and light-hole bands. The CdTe substrate is oriented along (001), and for this orientation the strain tensor components are given by
$\epsilon_{xx}=\epsilon_{yy}=\epsilon$, $\epsilon_{zz}=-2\epsilon C_{12}/C_{11}$ and $\epsilon_{i\neq j}=0$ for $i,j=x,y,z$.
Here, $C_{11}$ and $C_{12}$ are elastic stiffness constants. Because the off-diagonal components of the strain tensor
(the shear components) are zero, there are no internal electric fields generated in the HgTe layer due
to piezoelectric effects.
The strain tensor components for an arbitrary growth direction can be determined using the model of De~Caro
\textit{et al} \cite{decaro1995}. The effects of the strain tensor can be incorporated in the Kane model
through the Bir-Pikus Hamiltonian\cite{bir1974}, which can be easily obtained from the Kane Hamiltonian
with the substitution $k_{i}k_{j}\rightarrow \epsilon_{ij}$.
For the present case, one finds that lattice strain shifts the light-hole and heavy-hole band-edges at the $\Gamma$-point and leads to the opening of a gap between them which is given by $E_{g}=-b(\epsilon_{xx}+\epsilon_{yy}-2\epsilon_{zz})$,
where $b$ is the uniaxial deformation potential, which amounts to -1.5V for HgTe
\cite{Adachi09}. For $C_{12}/C_{11}=0.68$ \cite{barabash1999} we obtain $E_{g}\approx 22$ meV.
Using the above model, we have calculated the band structure of a 0.3 \% strained 70nm thick HgTe slab, yielding the dispersion shown in Fig.~\ref{HgTe}.
At $\Gamma$ point, the conduction band edge is around 11meV while the valence band edge around -11meV, and the direct gap at $\Gamma$ point
amounts to 22meV. Within the bulk gap, we find two special states ( shown in blue in Fig.1)
which touch the valence band states at the $\Gamma$ point, but merge with the conduction band for large
$k$. In the present system, the Dirac point of the surface state is buried deep within the heavy hole valence
band, about 70meV below the valence band edge for the HgTe-CdTe interface and 100meV for the HgTe-Vacuum surface,
as shown schematically by the red guiding lines in Fig. \ref{HgTe}.
 Therefore, these special states (shown in blue in Fig.1)  consist of the surface states originating from the Dirac type dispersion between $\Gamma_6$ electron
and $\Gamma_8$ light-hole bands hybridized with the $\Gamma_8$ heavy hole bands\cite{dai2008,chu2010}.

In actual devices,  positions of the Dirac points can vary from ones shown here. This is due to band bending from the Hartree potential, which has not been taken into account in the present calculations. Obviously, the existence of the surface states is not sensitive to the Hartree potential due to their topological nature.

\begin{figure}[htbp]
\centering
\includegraphics[width=0.45\textwidth]{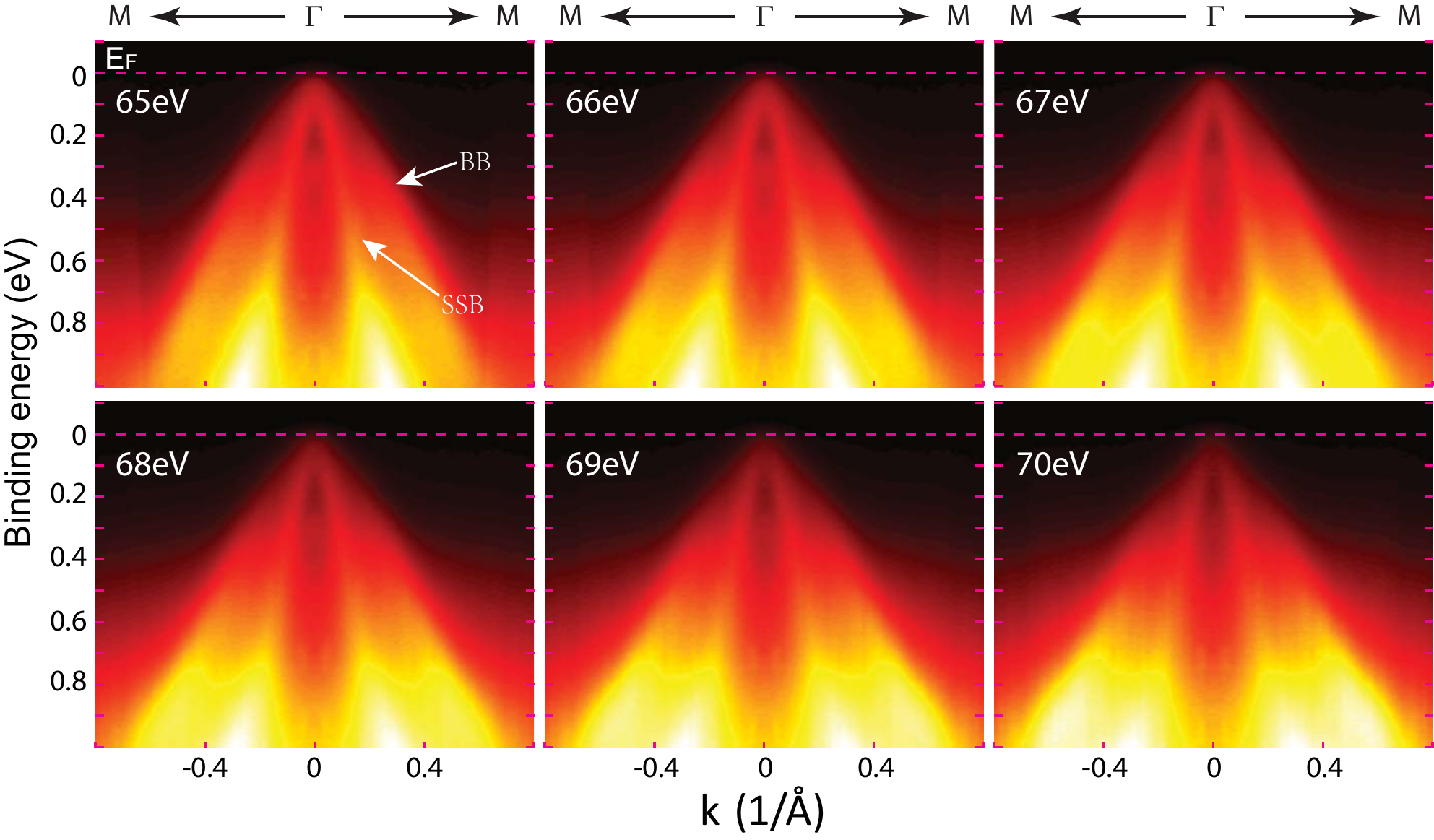}
\caption{{\bf a-e}, Energy dependent band structure along
M-$\Gamma$-M direction from 65-70eV excitation photons, with the
$E_F$, surface state band (SSB) and the bulk band (BB) marked.}
\label{ARPES}
\end{figure}

\section{Two Dirac cone model}
In this section, we will introduce an effective model with two Dirac cones
to describe the surface states of a slab of a topological insulator and investigate
the Landau level structure evolving from such a system when a perpendicular magnetic field is applied (this defines the z-direction).
On the surface of a strong topological insulator, the low-energy physics of the surface states can be
described by the Dirac Hamiltonian, with two components related to each other by time reversal,
which we refer to as spin for simplicity.
In our slab, we need to take two surfaces, which are perpendicular to the magnetic field and related to each other by
inversion, into account, and consequently, there are in total two Dirac cones and the
four basis states can be written as $|\alpha,\sigma\rangle$, where $\alpha=\pm$ denotes the upper
and lower surface and $\sigma=\uparrow,\downarrow$ denotes spin. Within this four state basis,
the effective Hamiltonian is written as
\begin{eqnarray}
    &&\hat{H}_{2D}=\hbar v_f\left(
    \begin{array}{cc}
        k_x\sigma_y-k_y\sigma_x&0\\
        0&-(k_x\sigma_y-k_y\sigma_x)
    \end{array}
    \right)\nonumber\\
    &&=\hbar v_f(k_x\sigma_y-k_y\sigma_x)\otimes
    \tau_z,
    \label{eq:H2D}
\end{eqnarray}
where the Pauli matrix $\sigma$ describes spin and $\tau$ the opposite surfaces;
the Hamiltonian is invariant under the inversion operation $P={\bf 1}\otimes\tau_x$
and the time reversal operation $\mathcal{T}=-i\sigma_y\otimes{\bf 1}K$
where $K$ indicates complex conjugation. We note that for zero gap HgTe/CdTe quantum wells\cite{buettner2010},
the effective Hamiltonian can also be described by two Dirac cones. However in that case,
the two Dirac cones are related to each other by time reversal, while here the two Dirac cones are related by inversion.
Besides the effective Hamiltonian $\hat{H}_{2D}$,
we also take into account two more terms that may be active in the sample:
an inversion breaking term\cite{shan2010}
\begin{eqnarray}
    \hat{H}_{ib}=\left(
    \begin{array}{cc}
        \Delta_i&0\\
        0&-\Delta_i
    \end{array}
    \right)=\Delta_i{\bf 1}\otimes \tau_z,
    \label{eq:Hib}
\end{eqnarray}
and a hybridization term between the two surface states, which should become important in very thin slabs:
\begin{eqnarray}
    \hat{H}_h=\left(
    \begin{array}{cc}
        0&\Delta_h\\
        \Delta_h&0
    \end{array}
    \right)=\Delta_h{\bf 1}\otimes\tau_x.
    \label{eq:Hh}
\end{eqnarray}
To study the effect of a magnetic field along the z-direction, we first need to make a Peierls substitution, i.e.
we replace ${\bf k}$ in the Hamiltonian $\hat{H}_{2D}$ with
$\pi={\bf k}+\frac{e}{\hbar}{\bf A}$ , where ${\bf A}=(0,B_0x,0)$
for a magnetic field ${\bf B}=B_0\hat{z}$. Additionally,
we need to consider a Zeeman type term, given by
\begin{eqnarray}
    \hat{H}_Z=\left(
    \begin{array}{cc}
        g^*\mu_BB_0\sigma_z&0\\
        0&g^*\mu_BB_0\sigma_z
    \end{array}
    \right)=g^*\mu_BB_0\sigma_z\otimes{\bf 1}
    \label{eq:HZ}
\end{eqnarray}
where $g^*$ is an effective g factor that takes orbital effects of nearby bands into account.

First let us have a look at the Landau levels of the Hamiltonian
$\hat{H}_{2D}$, which can be solved as
\begin{eqnarray}
    E_{\alpha t}(n)=t\sqrt{2ne\hbar v_f^2B_0}\qquad n=1,2,\cdots
    \label{eq:EigenEng1}
\end{eqnarray}
and $E_\alpha(n=0)=0$ for zero modes,
where $t=\pm$ denotes the electron and hole levels and $\alpha=\pm$
denotes the upper and lower surfaces. Obviously, each Landau level
is doubly degenerate since the the upper and lower surfaces are
identical and related by inversion. The zero modes will be half filled at charge neutrality,
and, thus due to the double degeneracy, the Hall plateaus
should appear only at odd number filling factors $n=1,3,5,\cdots$,
which, as described in the main article, is what we observe in low magnetic fields.

\begin{figure}[htbp]
\centering
\includegraphics[width=0.40\textwidth]{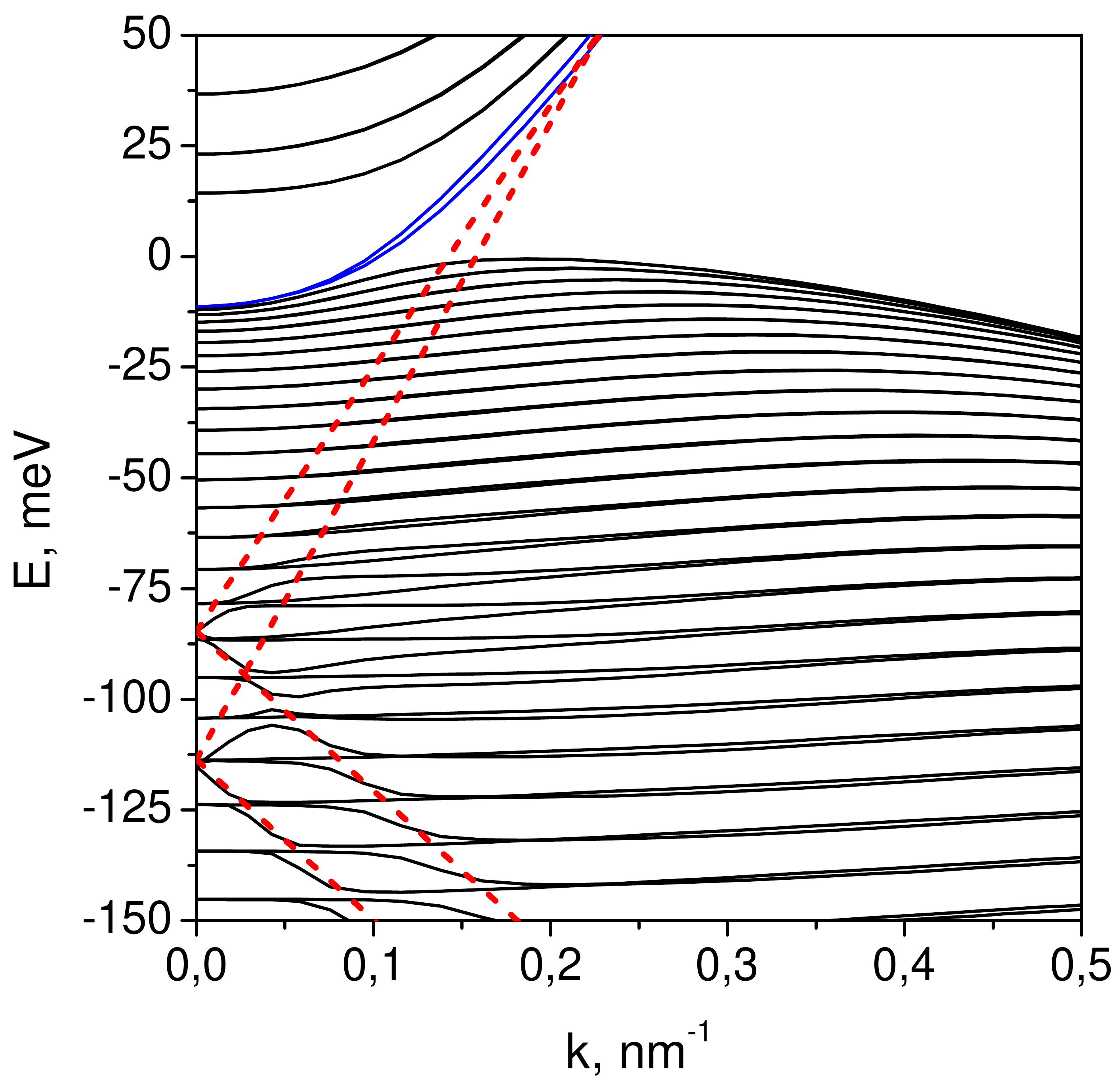}
\caption{Band structure of a 70nm thick 0.3 \% strained HgTe slab. The Dirac-like states in the gap are plotted in blue. The dashed red line schematically shows the dispersion of the Dirac surface states at the two opposite surfaces before their hybridization with the $\Gamma_8$ heavy hole bands. }
\label{HgTe}
\end{figure}

In high magnetic fields, Hall plateaus at even number filling factors
are also observed, which indicates a splitting of the Landau level degeneracy.
In zero gap HgTe/CdTe quantum wells (QWs) case\cite{buettner2010},
we have found that the even number filling factors observed in that system are caused by Zeeman splitting. It is thus
natural to first investigate the same mechanism for the present experiment. This implies solving the Landau level
problem for the Hamiltonian $\hat{H}_{2D}+\hat{H}_h+\hat{H}_Z$. One easily finds that in this case the Landau levels are given by
\begin{eqnarray}
    E_{\alpha t}(n)=t\sqrt{2ne\hbar v_f^2B_0+(|\mu_Bg^*B_0|+\alpha|\Delta_h|)^2}
    \label{eq:EigenEng2}
\end{eqnarray}
for $n=1,2,3,\cdots$ and $E_{\alpha}(0)=-g^*\mu_BB_0+\alpha|\Delta_h|$ for $n=0$.
From these expressions, we conclude that the double degeneracy remains when only the
Zeeman term is important but hybridization between the two surface states is negligible($\Delta_h=0$).
Therefore considerable hybridization between the two surface states (i.e. the slab has to be very thin) is
required in order to make this mechanism work. In HgTe, the decay length of the surface states
is estimated to be of order $2\sim 3$nm\cite{chu2010}, which is much smaller than the sample thickness of 70 nm.
It is thus unlikely that a Zeeman mechanism is responsible for the degeneracy lifting in the present experiment.

\begin{figure}[htbp]
\centering
\includegraphics[width=0.40\textwidth]{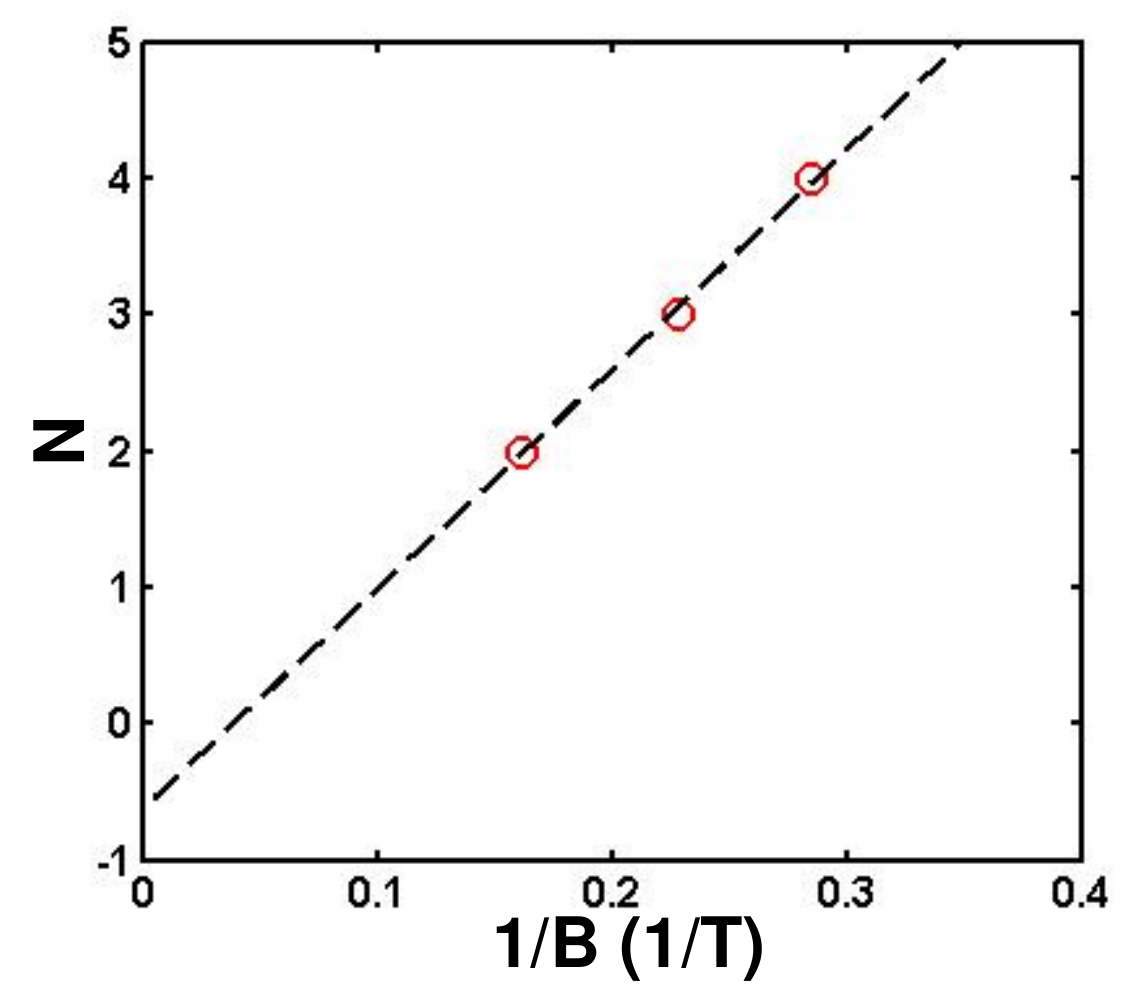}
\caption{ Landau level index for the data of fig. 2 in the main article plotted as a function of inverse magnetic field. }
\label{sigmaxy}
\end{figure}

The other possible mechanism for Landau level splitting is the breaking of
the inversion symmetry - which in practice occurs by the presence of the substrate. To see this, we solve the Landau level problem of the
Hamiltonian $\hat{H}_{2D}+\hat{H}_{ib}$ and obtain the eigen-energies as
\begin{eqnarray}
    E_{\alpha t}=t\sqrt{2ne\hbar v_f^2B_0}+\alpha\Delta_i
    \label{eq:EigenEng3}
\end{eqnarray}
for $n=1,2,3,\cdots$ and $E_\alpha(0)=\alpha\Delta_i$ for $n=0$.
From this expression, it is obvious that the degeneracy of the Landau levels is removed for any nonzero $\Delta_i$.
This makes inversion-symmetry breaking mechanism the most likely explanation for our observation of even filling factor at high magnetic fields.

For a comparison with the experimentally obtained Shubnikov-de Haas oscillations,
we calculate the density of states for the energy dispersion (\ref{eq:EigenEng3}) for two Dirac
cones with an inversion-symmetry breaking term. Taking $\hbar v_f=280meV\cdot nm$ and $\Delta_i=4.2meV$,
we find that the positions of the experimental Shubnikov-de Haas minima
can be fitted well with the densities of $n_1=3.7\times10^{11}cm^{-2}$ and
$n_2=4.8\times10^{11}cm^{-2}$ for the two Dirac cones, which are the traces shown in Fig. 3 of the main article.
Further evidence that this is the correct model for our observations comes from the odd filling factor ($\nu=9,7,5$)
quantum Hall plateaus we observe at low magnetic fields. A clear discrimination between the Landau level structure of a Dirac system and that of a conventional 2D electron gas can be obtained from plotting of the Landau level index as a function of 1/B\cite{novoselov2005,zhang2005}.
For a Dirac system with multiple Dirac cones, the filling factor is related to the Landau level index N by $\nu=m(N+1/2)$, where m is the number of the Dirac cones and equals to 2 here. Taking the magnetic field values
corresponding to the Hall plateaus $\nu=9,7,5$ from the $\sigma_{xy}$ curve in Fig. 2 in
the main article, and plotting N as a function of 1/B yields  Fig. \ref{sigmaxy}. The intercept of this plot for infinite magnetic field gives -1/2, which provides additional evidence
that the main physics of our system can be well described by the two Dirac cones model.

\bibliography{3DHgTe}

\end{document}